\documentstyle[aps,12pt]{revtex}
\begin{document}
\draft
\title{Phase randomness in a one-dimensional disordered absorbing medium}
\author{V. Freilikher and M. Pustilnik \cite{mp}}
\address{The Jack and Pearl Resnick Institute of Advanced Technology,\\
Department of Physics, Bar-Ilan University, Ramat-Gan 52900, Israel}
\date{\today }
\maketitle

\begin{abstract} 
Analytical study of the distribution of phase of the
transmission coefficient through 1D disordered absorbing system is
presented. The phase is shown to obey approximately Gaussian distribution.
An explicit expression for the variance is obtained, which shows that
absorption suppresses the fluctuations of the phase. The applicability of
the random phase approximation is discussed. 
\end{abstract}

\pacs{PACS numbers:  42.25.Bs, 72.15.Rn, 05.40.+j, 78.20.Ci} 

\newpage

Rigorous results for the reflection of waves from a disordered absorbing
one-dimensional medium have been obtained a long time ago \cite {Papa}.
Nevertheless, despite its evident significance, the problem of wave
propagation in absorbing disordered media has received less attention as
compared to the studies of waves in passive (nonabsorptive) systems. In
the pioneering work \cite{SJ} the possibility of the localization of light
has been predicted, which gave a new boost to the studies of classical
waves in random media. It was believed during the years that the effect of
absorption on the classical wave transport is similar to that of inelastic
scattering processes on electrons in disordered solids. Indeed, it is
well- known that when the incoherent effects are modelled by introducing
an optical (imaginary) potential, the theory simplifies drastically
\cite{SL}. Only recently it was understood that the role of absorption
differs fundamentally from the role of inelastic scattering for electrons
\cite{RW} (see also discussion in \cite{RK}, \cite{Y}). The source of this
difference is that absorption corresponds to the actual removal of
particles, but does not affect the phase coherence. Conversely, the
inelastic scattering causes the loss of the phase memory and suppresses
the interference effects, but preserves the number of particles. Recently
there have been a number of works dealing with the effect of absorption
(and related problem of the coherent amplification) on the wave transport
in random media \cite{RW}-\cite {we2}. 

We study in this paper the probability distribution of the phase $\theta $
of the transmission coefficient $t$ through 1D random absorbing media. It
is well-established now that in a passive (nonabsorbing) 1D medium the
interference leads to the strong localization of all eigenstates in the
presence of arbitrary weak disorder. If the length of the scattering
region $L$ is large compared to the characteristic length of elastic
scattering $l$, the transmittance $T=\left| t\right|^2$ decreases
exponentially with the length in average. On the other hand, $T$ strongly
fluctuates and obeys the log-normal distribution. If the disorder is weak,
the phases of the reflection and transmission coefficients become random
\cite{SAJ}, \cite{P}.  One may assume that the phase $\varphi$ of the
reflection coefficient is distributed uniformly in the interval $\left(
0,2\pi \right)$ and independently on the amplitude of the reflection 
coefficient and amplitude and phase of the transmission coefficient. This
assumption will be referred below as the random phase approximation (RPA).
RPA, however, does not implies that the phase $\theta$ of the transmission
coefficient is also uniformly distributed. Indeed, the recently obtained 
formula reads \cite{MT}: 
\begin{equation}
\left\langle t/t^{*}\right\rangle =\left\langle e^{2i\theta }\right\rangle
=\left( 1+L/l\right) e^{-3L/l}.  \label{mello}
\end{equation}
Note that calculating this average with the uniform distribution for $\theta $
one would obtain zero. 

The arguments presented above suggest that RPA can be used for calculation 
of the distribution of the absolute phase 
$\theta \in \left( -\infty ,\infty
\right)$. When the absorption is zero, $\theta$ relates to the density of 
states in the scatterer by $\rho =\pi
^{-1}\left[ \partial (kL+\theta) /\partial E\right] $, $E=k^2$ \cite{Th}. 
Therefore 
$\left(k+\theta /L \right)/ \pi $
 is the total number of states per unit length, which 
is self-averaged in the limit $L/l\gg 1$ \cite{LGP}, and obeys almost
Gaussian distribution with 
$\mathop{\rm var}\left( \theta /L\right) \propto L^{-1}$. 
The natural question arises in the
context of the discussion above is how absorption affects this
distribution. This question is addressed below.

A wave $U(z)$ with the wavenumber $k$ propagating in a one dimensional 
random medium satisfies the
equation 
\begin{equation}
\left( \frac{d^2}{dz^2}+k^2\epsilon \left( z\right) \right)
U(z)=0,  \label{Helmholtz}
\end{equation}
where the dielectric permittivity $\epsilon \left( z\right)$ is equal 
to $1$ for $z<0$ and $z>L$ and 
$\epsilon \left( z\right)=1+\delta\epsilon \left( z\right)$ 
for $0\leq z\leq L$. 
We assume for simplicity that the imaginary part of 
$\delta\epsilon \left( z\right)$, which is responsible for the absorption, 
is constant, while the real part $\varepsilon \left( z\right)$ is a 
random function of $z$, which may be approximated by Gaussian white noise:
\begin{equation}
\delta\epsilon \left( z\right)=\varepsilon \left( z\right) +i/kl_a,
\end{equation}
\begin{equation}
\left\langle \varepsilon(z) \right\rangle =0,\;\;
\left\langle \varepsilon(z) \varepsilon( z^{\prime}) \right\rangle =
\frac 4 {k^2l} \delta \left( z-z^{\prime }\right).
\end{equation}                                            
Here the length of elastic scattering $l$ in a system without absorption
and the length of absorption $l_a$ in a system without disorder have been 
introduced.
The reflection ($r$) and transmission ($t$) coefficients are defined 
according to the wave behavior outside the disordered region: 
\[
U\propto \left\{ 
\begin{array}{ll}
e^{-ikz}+re^{ikz}, & z\geq L \\ 
te^{-ikz}, & z\leq 0
\end{array}
\right.
\]

To proceed, we introduce the joint distribution function 
\begin{equation}
{\cal P}_L\left( x,\theta \right) =\left\langle \delta \left( x-x\left(
L\right) \right) \delta \left( \theta -\theta \left( L\right) \right)
\right\rangle ,  \label{def}
\end{equation}
where we have used the parametrization $r=\tanh \left( x/2\right)
e^{i\varphi },\;t=\left| t\right| e^{i\theta }$, $-\infty <\theta <\infty $. 
To derive the Fokker-Planck equation for ${\cal P}_L\left( x,\theta
\right)$ it is convenient to use the so-called invariant imbedding
equations for $r$ and $t$ (a detailed description of the similar
calculation can be found for instance in \cite{We}, \cite{RD}). Assuming 
that RPA is 
applicable for length $L$ larger than the wavelength ($kL\gg 1$) and in 
the limit of weak scattering ($kl \gg 1$) and weak absorption ($kl_a 
\gg 1$), we obtain the equation
\begin{equation}
\frac \partial {\partial S}{\cal P}_S\left( x,\theta \right) =\left( \frac 
\partial {\partial x}\left( \frac \partial {\partial x}+\frac{d\Omega _\beta 
}{dx}\right) +F\left( x\right) \frac{\partial ^2}{\partial \theta ^2}\right) 
{\cal P}_S\left( x,\theta \right) ,  \label{FP}
\end{equation}
where 
\begin{equation}
\Omega _\beta \left( x\right) =-\ln \left( \sinh x\right) +\beta \cosh x,
\label{omega}
\end{equation}
\[
F\left( x\right) =\frac 14\left( 3-\frac 1{\cosh ^2x/2}\right) =\frac 14%
\left( 2+R_S\right) . 
\]
Here $S=L/l$ and $l$ is the scattering length for zero absorption. The 
parameter $\beta =l/l_a$ describes the relative strength of absorption. 
Alternatively, Eq.(\ref{FP}) can be derived by the method of Ref. 
\cite{MT}. Indeed, for $\beta =0$ Eq.(\ref{FP}) is equivalent to the equation
for expectation values of Ref. \cite{M}. It should be noticed that it is 
not necessary to assume the white noise distribution for 
$ \varepsilon(z)$ to derive Eq.(\ref{FP}). It is enough to require the 
weakness of scattering $kl \gg 1$ \cite{MT}.

After integration over $\theta $ Eq. (\ref{FP}) reduces to the well-known 
equation for the probability density distribution of $x$ \cite{Papa}. 
The solution of the later equation saturates at $S\gg 4$ to the limiting $S$-independent 
distribution  \cite{Papa}, \cite{We}, \cite{K} 
\begin{equation}
{\cal P}_\infty \left( x\right) =\beta e^{\beta -\Omega _\beta \left(
x\right) }  \label{stat}
\end{equation}
with $\Omega _\beta $ given by Eq. (\ref{omega}). 
The distribution (\ref{stat}) implies the following limiting value for 
the average reflectance \cite{Papa}:
\begin{equation}
R_\infty =\lim_{S\rightarrow \infty }\left\langle R_S\right\rangle =1+2\beta e^{2\beta }\mathop{\rm Ei}
\left( -2\beta \right) \stackrel{}{\simeq }\left\{ 
\begin{array}{ll}
1-2\beta \ln \left( 1/\beta \right) , & \beta \ll 1 \\ 
\left( 2\beta \right) ^{-1}, & \beta \gg 1
\end{array}
\right.  \label{RR}
\end{equation}
where ${\rm Ei}(x)=\int\limits_{-\infty}^x dye^y/y$ .

Here is a point to discuss the validity of RPA, which was essentiall for 
derivation of Eq.(\ref{FP}), and therefore underlies the result of 
Eq.(\ref{RR}) that predicts a 
monotonic decrese of $R_\infty$ with $\beta$. Despite this result being 
known for more than twenty years, it was only recently understood \cite{RK}, 
\cite{Y} that this prediction becomes wrong for large $\beta $. Moreover, 
it has been verified numerically that RPA fails at the same values 
of $\beta$ \cite{Y}. Let us estimate now the range of validity of RPA.  In 
the limit $\beta=l/l_a\gg 1$ the wave penetrates into disordered absorbing 
medium at a distance which is much smaller than elastic scattering length $l$.
This means that disorder may lead only to a small correction to $R_\infty $, 
which can
be taken into account by means of perturbation expansion in the small
parameter $1/\beta $: 
\begin{equation}
\text{ }R_\infty =R^0+R^1\beta ^{-1}+O\left( \beta ^{-2}\right) . \label{RRR}
\end{equation}
The zeroth-order term here does not depend on $\beta $ but
on the parameter $kl_a$ only. In the limit $kl_a\ll 1$ (strong absorption, 
which was the case in \cite{RK}) this term is equal to $1$. This is easy to understand since this limit
corresponds to a large conductivity, where the skin-effect results in the 
almost perfect reflection. However, we are considering here the opposite limit
of the weak absorption ($kl_a\gg 1$). In this limit, $\beta \gg 1$
corresponds to the absence of the elastic scattering rather than to the strong
absorption, and $R^0$ is nothing but the reflectance of the ideal (ordered) 
lossy system. Calculating the reflectance $R_\infty $ in the
lowest order on $\left( kl_a\right) ^{-1}$ and $\beta ^{-1}$, we find that
Eq.(\ref{RRR}) takes the form 
\begin{equation}
R_\infty \simeq \left( 4kl_a\right) ^{-2}\text{ }+\left( 2\beta \right)
^{-1}\left[ 1+O\left( 1/kl\right) \right] .  \label{RRRR}
\end{equation}
Eq.(\ref{RRRR}) is exact for $kl_a\gg 1$ and $kl\gg 1$. 
The first term in the r.h.s. in Eq.(\ref{RRRR}) is the reflectance of the 
semi-infinite clean media with the wave number 
$k_1=\left[ k^2\left( 1+i/kl_a\right) \right] ^{1/2}\simeq k\left
( 1+i/2kl_a\right)$ and is absent in Eq.(\ref{RR}). 
The second term represents the correction due to disorder and for $\beta\gg 1$
coincides exactly with Eq.(\ref{RR}).  Since RPA was the only approximation
made in derivation of the non-perturbative result Eq.(\ref{RR}), one 
immediately finds that RPA is applicable if, in addition to the weakness of both the scattering and absorption, the following inequality holds: 
$\left( 4kl_a\right) ^{-2}\ll \left( 2\beta \right) ^{-1}$ , or
\begin{equation}
\beta \ll \beta _0\sim 8\left( kl_a\right) ^2.  \label{limit}
\end{equation}
These simple arguments are consistent with the recent 
numerical simulations \cite
{Y}, which clearly show that RPA is valid only when $\beta $ is smaller than
some $\overline{\beta }$ $\left( \beta \ll \overline{\beta }\right) $, for
which $R_\infty \left( \beta \right) $ takes its minimum. Eq. (\ref{limit})
provides therefore the value of $\overline{\beta }\sim \beta _0$ in the
leading order in the parameter $1/kl$, characterizing the weakness of
disorder.
 
Now that we have established the limits of the validity of Eq.(\ref{FP}) 
in Eq.(\ref{limit}), let us turn to its solution. 
It is natural to assume 
that correlations between the phase $\theta $ and reflectance $R$
are negligible, and distribution (\ref{def}) factorizes (see e.g. \cite
{Mello}): 
\begin{equation}
{\cal P}_S\left( x,\theta \right) \simeq P_S\left( x\right) P_S\left( \theta
\right) ,\;S\gg 1  \label{con}
\end{equation}
Integrating Eq. (\ref{FP}) by making use of this conjecture, we get
\begin{equation}
\frac \partial {\partial S}{\cal P}_S\left( \theta \right) \simeq \frac 14%
\left( 2+\left\langle R_S\right\rangle \right) \frac{\partial ^2}{\partial
\theta ^2}{\cal P}_S\left( \theta \right) ,  \label{fptheta}
\end{equation}
which immediately results in the Gaussian distribution with zero average $%
\left\langle \theta \right\rangle =0$ and variance 
\begin{equation}
\mathop{\rm var}
\theta =\left( 1+R_\infty /2\right) S.  \label{varb}
\end{equation}
One can see from Eqs.(\ref{RR}),(\ref{varb}) that the absorption 
suppresses fluctuations of $\theta $. Combining Eq.(\ref{varb}) with the 
recently obtained result ${\rm var}(\ln T)=2R_\infty S$  \cite{We} we 
obtain another useful formula 
\begin{equation}
\mathop{\rm var}
\theta =S+\frac 14%
\mathop{\rm var}
\left( \ln T\right) . \label{theta-lnT}
\end{equation}
Calculating $\left\langle e^{2i\theta }\right\rangle $ for $\beta =0$ with
the Gaussian distribution specified by (\ref{varb}) one recovers Eq.(\ref
{mello}) with an exponential accuracy.

It is interesting to note that Eq.(\ref{varb}) is rather general, and can
be obtained without the conjecture (\ref{con}). To do this we use the
formula 
\begin{equation}
{\rm var}\theta =-\left( \frac{d^2\left\langle e^{ip\theta }\right\rangle }{%
d\left( p^2\right) }\right) _{p=0} . \label{dddd}
\end{equation}
To calculate the characteristic function $\left\langle e^{ip\theta
}\right\rangle $ we consider the Fourier-transform of (\ref{FP}) 
\begin{equation}
\frac \partial {\partial S}P_S\left( x;p\right) =\left( \frac \partial {%
\partial x}\left( \frac \partial {\partial x}+\frac{d\Omega _\beta }{dx}%
\right) -p^2F\left( x\right) \right) P_S\left( x;p\right) ,  \label{FPP}
\end{equation}
Integrated over $x$, $P_S\left( x;p\right) $ gives us a characteristic
function $\left\langle e^{ip\theta }\right\rangle $. Following the procedure
used in \cite{We} we substitute 
\[
P_S\left( x;p\right) =e^{-\Omega _\beta /2}\Psi(x;S) 
\]
which yields an imaginary time Scr\"odinger equation for $\Psi $. Solution
of this equation in spectral representation takes the form 
\[
\Psi \left( x;S\right) =\sum_ne^{-E_nS}\psi _n\left( x\right) \left[ \psi
_n\left( y\right) e^{\Omega _\beta \left( y\right) /2}\right] _{y\rightarrow
0} , 
\]
where $E_n$ and $\psi _n$ are eigenvalues and real normalized eigenfunctions
of the Scr\"odinger operator 
\[
\left( -\frac{d^2}{dx^2}+V_p\left( x\right) \right) \psi _n\left( x\right)
=E_n\psi _n\left( x\right) 
\]
with potential 
\begin{equation}
V_p\left( x\right) =\frac 14\left( 1-\sinh ^{-2}x\right) +\frac{p^2}4\left(
3-\cosh ^{-2}x/2\right) -\beta \cosh x+\frac{\beta ^2}4\sinh ^2x  \label{po}
\end{equation}
It can be easily seen \cite{We} that the potential $V_p\left( x\right) $ has a minimum at 
$x\sim \ln \left( 4/\beta \right) $, and the large $S$ behavior of $%
P_S\left( x;p\right) $ is dominated by the energy level localized in this
well. For $p=0$ the energy of this level is exactly $0$ and the
corresponding wave function is $\psi _0^0=\left( {\cal P}_\infty \left(
x\right) \right) ^{1/2}=\beta ^{1/2}e^{\beta /2-\Omega _\beta \left(
x\right) }$ . For small $p^2$ (but for arbitrary $\beta $) the perturbation
theory yields: 
\[
E_0\left( p\right) =\left\langle \psi _0^0\right| p^2F\left( x\right) \left|
\psi _0^0\right\rangle +o\left( p^2\right) =\frac 14p^2\left( 2+R_\infty
\right) +o\left( p^2\right) 
\]
Therefore for $S\gg 4$ 
\[
\left\langle e^{ip\theta }\right\rangle \propto e^{-E_0S}=\exp \left[ -\frac 
14p^2\left( 2+R_\infty \right) +o\left( p^2\right) \right] .
\]
Making use of Eq. (\ref{dddd}), we end up with Eq. (\ref{varb}).

To conclude, we have studied analytically the influence of small absorption
on the distribution of the phases of the reflection and transmission 
coefficients through 1D weakly disordered random media. It was shown that the phase $\varphi$ of the reflection coefficient can be treated as a uniformly distributed when the absorption is weak enough to satisfy the inequality Eq.(\ref{limit}). An explicit expression for 
the variance of the phase $\theta$ of the transmission coefficient was obtained (Eq.(\ref{varb})). The result shows that the presence of the absorption slows down the evolution of $\theta$ towards a random limit. 

We thank I. Yurkevich for helpful discussions.


\begin{references}
\bibitem[*]{mp}  
present address: Department of Physics,
University of Illinois at Chicago, Chicago, IL 60607

\bibitem{Papa}  W. Kohler and G. Papanicolaou, SIAM J. Appl. Math. {\bf 30},
263 (1976).

\bibitem{SJ}  S. John, Phys. Rev. Lett. {\bf 53}, 2169 (1984).

\bibitem{SL}  A. D. Stone and P. A. Lee, Phys. Rev. Lett. {\bf 54}, 1196
(1985).

\bibitem{RW}  R. L. Weaver, Phys. Rev. B{\bf \ 47}, 1077 (1993); M. Yosefin,
Europhys. Lett. {\bf 25}, 675 (1994); M. Kaveh and E. Kogan, in {\it %
Photonic Band Gaps and Localization}, ed. by C. Soukoulis (Plenum Press, New
York, 1993), p. 187.

\bibitem{RK}  A. Rubio and N. Kumar, Phys. Rev. B {\bf 47}, 2420 (1993).

\bibitem{Y}  A. K. Gupta and A. M. Jayannavar, Phys. Rev. B{\bf \ 52}, 4156
(1995).

\bibitem{We}  V. Freilikher, M. Pustilnik, and I. Yurkevich, Phys. Rev.
Lett. {\bf 73}, 810 (1994); Phys. Rev. B {\bf 50}, 6017 (1994).

\bibitem{K}  P. Pradhan and N. Kumar, Phys. Rev. B {\bf 50}, 9644 (1994).

\bibitem{Z}  A. Yu. Zyuzin, Phys. Rev. E {\bf 51}, 5274 (1995).

\bibitem{Beenakker1}  C. W. J. Beenakker, J. C. J. Paasschens, and P. W.
Brouwer, preprint (cond-mat/9601024); T. Sh. Misirpashaev and C. W. J.
Beenakker, preprint (cond-mat/9607119).

\bibitem{Beenakker2}  J. C. J. Paasschens, T. Sh. Misirpashaev and C. W. J.
Beenakker, preprint (cond-mat/9602048); T. Sh. Misirpashaev, J. C. J.
Paasschens and C. W. J. Beenakker, preprint (cond-mat/9607118).

\bibitem{we2}  V. Freilikher, M. Pustilnik, and I. Yurkevich, preprint
(cond-mat/9605090).

\bibitem{SAJ}  A. D. Stone, D. C. Allan and J. D. Joannopoulos, Phys. Rev. B 
{\bf 27}, 836 (1983).

\bibitem{P}  J. B. Pendry, Adv. in Phys. {\bf 43}, 461 (1994), P. D. Kirkman
and J. B. Pendry, J. Phys. C {\bf 17}, 4327 (1984).

\bibitem{MT}  P. A. Mello and S. Tomsovic, Phys. Rev. Lett. {\bf 67}, 342
(1991).

\bibitem{Th}  D. J. Thouless, J. Phys. C {\bf 5}, 77 (1972); P. W. Anderson
and P. A. Lee, Progr. Theor. Phys. Suppl. {\bf 69}, 212 (1980).

\bibitem{LGP}  I. M. Lifshits, S. A. Gredeskul and L. A. Pastur, {\it %
Introduction to the Theory of Disordered Systems} (Wiley, New York, 1988).

\bibitem{RD}  R. Rammal and B. Doucot, J. Physique {\bf 48}, 509 (1987).

\bibitem{M}  P. A. Mello, Phys. Rev. B {\bf 47}, 16358 (1993).

\bibitem{Mello}  P. A. Mello, Phys. Rev. B {\bf 35}, 1082 (1987).
\end{references}
\end{document}